\documentclass[12pt,twoside]{article}
\usepackage{fleqn,espcrc1,epsfig}

\begin{document}

\title{Quasideuteron Configurations  in $^{46}$V and $^{58}$Cu} 
          
\author{P. von Brentano\address{ Institut f\"ur Kernphysik, Universit\"at 
zu K\"oln, 50937 K\"oln, Germany}
A.~F.~Lisetskiy$\,^a$, 
A.~Dewald$\,^a$, 
C.~Frie{\ss}ner$\,^a$, A.~Schmidt$\,^a$, I.~Schneider$\,^a$, 
 N.~Pietralla$\,^{a,}$\address{ Wright Nuclear Structure Laboratory, 
Yale University, New Haven, \\
Connecticut 06520-8124, USA}}

\maketitle

\begin{abstract}
  The data on low spin states in the odd-odd nuclei $^{46}$V and $^{58}$Cu 
  investigated  with the $^{46}$Ti\,(p,n$\gamma$)$^{46}$V,
  $^{32}$S\,($^{16}$O,pn)$^{46}$V and $^{58}$Ni\,(p,n$\gamma$)$^{58}$Cu
  reactions at the FN-{\sc Tandem} accelerator in Cologne are reported. 
  The states containing large {\em quasideuteron} components are identified 
  from the strong isovector M1 transitions, from shell model 
  calculations and from experimental data for low-lying states.  
\end{abstract}

\section{INTRODUCTION}

\begin{figure}[t]
\epsfxsize 11.0cm \centerline{\epsfbox{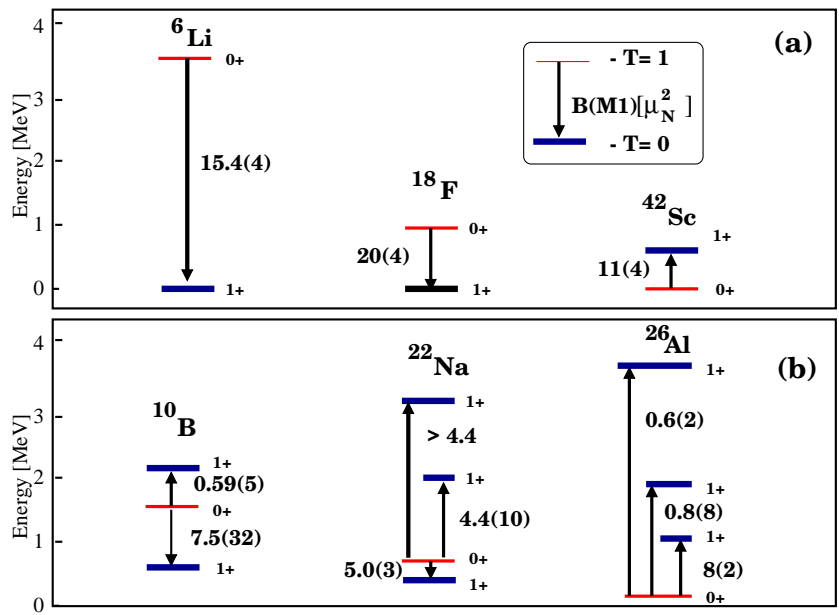}}
\caption{Strong isovector M1 $0^+ \rightarrow 1^+$ transitions in odd-odd 
N=Z nuclei. Panel (a) shows experimental data for the nuclei where the 
strength is concentrated in one $0^+_1 \rightarrow 1^+_1$ transition. 
Panel (b) gives examples of fragmentation of quasideuteron M1 strength. 
The experimental data are  taken from \cite{NS}.}
\label{fig:strong}
\end{figure}

An understanding of the nuclear phenomena related to the proton-neutron 
interaction is a cornerstone of many contemporary investigations in  
nuclear structure physics \cite{Lenzi99,OLeary99,Fries98,Schneider99,Schmidt,Sve98,Exp1,Grz,Vin98,Vogel,Mac2,Faes99,Isa97,Buc97,Ots98,Lis99}.
An important laboratory to study the effects coming from the pn 
interaction in the T=0 channel and its competition with the T=1 pn channel  
are odd-odd N=Z nuclei where the low-lying T=0 and T=1 states are 
almost degenerate. One of the interesting phenomenon related to the 
interplay between low-lying T=0 and T=1 states is the occurrence of very 
strong $\Delta$T=1 M1 transitions between them in some odd-odd N=Z nuclei 
(see Fig \ref{fig:strong}). 
The positive 
interference of orbital and large spin parts of reduced $\Delta$T=1 M1 
matrix elements between the states formed by the odd proton and the odd 
neutron 
occupying single $j=l+1/2$ orbitals explains the enhancement of M1 
transitions in odd-odd N=Z nuclei \cite{Lis99}. 
In other cases ($j=l-1/2$) the M1 strength almost vanishes due to the 
destructive interference of orbital and spin parts. 
The states having one proton one 
neutron ($\pi j \times \nu j)_{J,T}$ structure with $j=l+1/2$ contain 
a large component with total orbital angular momentum $L=J-1$
similarly to the ground state of real deuteron : $J^\pi=1^+$, $L=0$. 
Moreover the theoretical B(M1;$0^+ \rightarrow 1^+$) value 
for the hypothetical deuteron with a bound $J^\pi=0^+$, T=1 state would be very 
large and amounts to 16 $\mu_N^2$. Therefore we propose to call one proton 
one neutron 
configurations in a single $j=l+1/2$ orbital  {\em quasideuteron} 
configurations and to consider strong M1 transitions between the states 
of this structure as an indication of deuteron-like correlations. 

In light and medium-heavy nuclei the $j=N+1/2$ orbital (N is the principal 
quantum number) is well separated from other spherical
orbitals. In this case quasideuteron configurations are weakly mixed with 
other configurations resulting in very strong M1 $0^+_1\rightarrow1^+_1$ 
transitions in $^6$Li, $^{18}$F and $^{42}$Sc ( see Fig. \ref{fig:strong}a).  
An experimental indication of comparably strong M1 transition 
was also found recently in $^{54}$Co \cite{Schneider99}.

For larger number of valence protons and neutrons in a single 
$j=N+1/2$ orbital in odd-odd N=Z nuclei the quasideuteron 
configurations are fragmented among two or three states due to the 
collective effects \cite{Lis00}. 
This is actually observed in $^{10}$B,$^{22}$Na and $^{26}$Al nuclei 
( see Fig. \ref{fig:strong}b). In the case when $j=l+1/2$ and $l=N-2$, the 
quasideuteron  configurations are strongly fragmented due to the mixing 
with the configurations which involve other closely lying orbitals.

In the present work we would like to illustrate both effects 
(collectivity and nearness of $j=l-1/2$ orbitals) causing the fragmentation 
of the quasideuteron configurations using experimental data for odd-odd N=Z 
nuclei $^{46}$V and $^{58}$Cu.    
       
\section{LOW-SPIN BAND STRUCTURE OF $^{46}$V}
Recently the low spin structure of odd-odd N=Z nucleus 
$^{46}$V  was studied in Cologne \cite{Fries98,Friesdoc,Schneiderdoc}.
Parallel to this work there were two studies of high spin states performed
 by C.D.~O'Leary {\it et.al.} \cite{OLeary99} and by S.M.Lenzi {\it et.al.} 
\cite{Lenzi99}.

Low-spin states of $^{46}$V were populated using fusion 
evaporation $^{46}$Ti(p,\,n$\gamma$) $^{46}$V \cite{Fries98,Friesdoc} and 
$^{32}$S($^{16}$O,\,pn)$^{46}$V reactions \cite{Schneiderdoc}. The beam was 
delivered by the FN-TANDEM accelerator of the University of Cologne.  
In total, seven new spin assignments and five new parity assignments 
were made \cite{Fries98} as well as four lifetimes were measured 
\cite{Friesdoc,Schneiderdoc}. Using new experimental data we were 
able to extract the absolute B(M1) and B(E2) values or to determine their 
lower limits for 12 transitions between negative and positive parity 
states ( some of them are shown in Table \ref{tabvgl}).  

\begin{figure}[t]
\vspace{-0cm}
\epsfxsize 10.5cm \centerline{\epsfbox{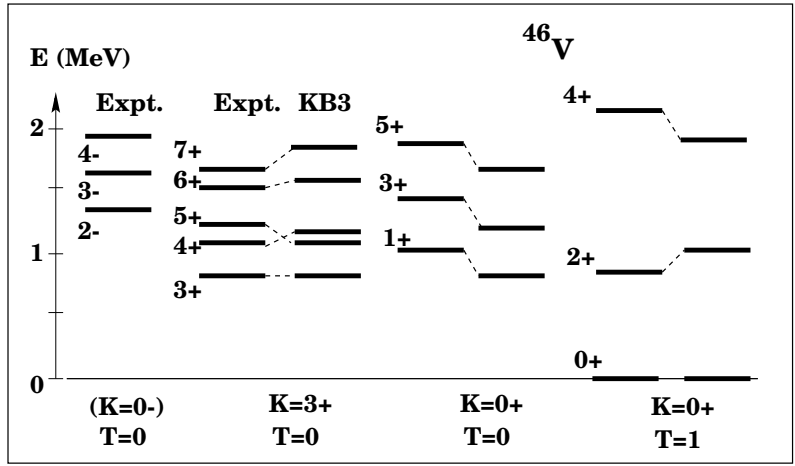}}
\vspace{-0cm}
\caption{ Band classification of the low-lying states
 in $^{46}$V . Experimental (left column of the band) and 
shell model (right column of the band) energies of the levels with 
certain spin and parity quantum numbers are shown. Shell model results 
are taken from \protect\cite{Fries98}. }
\label{band46}
\end{figure}

The experimental data were compared to shell model  calculations 
of the positive parity states of $^{46}$V in the full pf-shell without 
truncation with KB3 and FPD6 residual interactions. The calculations were 
done by the Tokyo group \cite{Fries98}.  
\begin{table}[t]
\begin{center}
\caption{The experimental and calculated $B(M1,J_i^+ \rightarrow J_f^+)$ and 
 $B(E2,J_i^+ \rightarrow J_f^+)$ values. The KB3 shell model calculations 
 are taken from \protect \cite{Fries98}.}
\label{tabvgl}
\begin{tabular}{cccccc}
\\
 \multicolumn{6}{c}{$^{46}V$} \\
\hline
Transition & \multicolumn{2}{c}{$B(M1;J^\pi_i \rightarrow J^\pi_f)$}& 
Transition & \multicolumn{2}{c}{$B(E2;J^\pi_i \rightarrow J^\pi_f)$}  \\
\hline 
$(J^\pi_i,K_i) \rightarrow (J^\pi_f,K_f)$ & Expt. & KB3 &
$(J^\pi_i,K_i) \rightarrow (J^\pi_f,K_f)$ & Expt. & KB3  \\
\hline
$\Delta$T=1 & \multicolumn{2}{c}{$\mu_N^2$} & 
$\Delta$T=0 & \multicolumn{2}{c}{e$^2$fm$^4$} \\ 
\hline
$({0_1}^{+},0) \rightarrow ({1_1}^{+},0)$ & $\ge$2.31 & 3.80 &
$({2_1}^{+},0) \rightarrow ({0_1}^{+},0)$ & 137(35)   & 143   \\
$({3_2}^{+},0) \rightarrow ({2_1}^{+},0)$ & 1.98(71) & 1.25 &  
$({4_2}^{+},0) \rightarrow ({2_1}^{+},0)$ & $\ge$169 & 187  \\
$({4_2}^{+},0) \rightarrow ({3_1}^{+},3)$ & 0.012(3) & 0.08 &  
$({4_1}^{+},3) \rightarrow ({3_1}^{+},3)$ & 200(50)  & 234  \\
$({4_2}^{+},0) \rightarrow ({3_2}^{+},0)$ & 0.57(15) & 0.85 & 
$({5_1}^{+},3) \rightarrow ({3_1}^{+},3)$ & 66(14) &  65 \\
$({4_2}^{+},0) \rightarrow ({5_1}^{+},3)$ & 0.02(1) & 0.02  
& & &\\
$({4_2}^{+},0) \rightarrow ({5_2}^{+},0)$ & 0.55(13) & 1.75 
& & &\\
\hline
\end{tabular}
\end{center}
\end{table}
Most of the experimental data are well reproduced in shell model 
calculations \cite{Fries98} (see Table \ref{tabvgl} and Fig. \ref{band46}).  
Experimental and shell model results can be interpreted also in terms of 
the Nilsson scheme. According to this model the odd proton and the odd 
neutron in 
$^{46}$V should occupy the Nilsson [321]$\Omega$=$3/2^-$ orbital.  Then the 
low-lying states in $^{46}$V should form K=0,T=1 (even spins), 
K=0,T=0 (odd spins) and K=3, T=0 bands. An analysis of experimental 
and theoretical B(E2) values shows that such a classification of low-spin 
states in $^{46}$V is possible ( see Fig. \ref{band46}).   
Based on the B(E2;$2^+_1 \rightarrow 0^+_1$) value one can show that the 
deformation of the K=0,T=1 band corresponds to $\beta \approx$0.23. Likewise  
B(E2;$4^+_1 \rightarrow 3^+_1$) and B(E2;$5^+_1 \rightarrow 3^+_1$) values 
help to estimate the value of deformation parameter for the K=3,T=0 band: 
$\beta \approx 0.23$, which is exactly the same as for the K=0,T=1 band. 

Having observed collective features in the structure of  $^{46}$V 
we identified also strong M1 transitions (see Table \ref{tabvgl}) which have 
non-collective quasideuteron nature and could be described quantitatively in 
frames of the Nilsson model too \cite{Lis00}. According to the quasideuteron 
picture \cite{Lis99} (see also \cite{Zamick}) one should expect a very 
strong M1 $0^+ \rightarrow 1^+$ transition ( 18 $\mu_N^2$) for odd-odd N=Z 
nuclei in the $f_{7/2}$ orbital. The main part of 
this strength is predicted by Nilsson model and shell model to be distributed 
among three $0^+_1 \rightarrow 1^+_i$ transitions in $^{46}$V nucleus.
 Supposing that theoretical and experimental ratios of 
B(M1;$3^+_2\rightarrow 2^+_1$) and B(M1;$0^+_1\rightarrow 1^+_1$) values are 
 similar, one can actually estimate that a large part of this strength 
[6(2)$\mu_N^2$] is concentrated in the $0^+_1 \rightarrow 1^+_1$ transition.

Furthermore it follows from large scale shell model calculations for 
$^{46}$V with KB3 and FPD6 residual interactions that the ratios of 
B(E2;$4^+_2\rightarrow 2^+_1$) and B(E2;$2^+_1\rightarrow 0^+_1$) values
 are 1.39 and 1.31 for FPD6 and KB3 interactions, respectively, i.e. the ratio 
is just slightly model dependent. From our measured lifetime of 
the $2^+_1,T=1$ state and the
extracted B(E2;$2^+_1\rightarrow 0^+_1$) value  one can expect that 
B(E2;$4^+_2\rightarrow 2^+_1$)= 1.35B(E2;$2^+_1\rightarrow 0^+_1$)= 185(47) 
e$^2$fm$^4$. This number is in a good agreement with the lower limit of 
169 e$^2$fm$^4$ which we have obtained from our new experimental data.
The $4^+_2,T=1$ state strongly decays also to some T=0 states and we 
know the intensity and multipole mixing ratios for these transitions. Using 
these experimental data we obtain the absolute strength of  four M1 
transitions 
(see Table \ref{tabvgl}). This example clearly shows that 
some of the $\Delta$T=1 M1 transitions are retarded due to the K quantum 
number selection rule ($\Delta$K=3 M1 transitions are forbidden) while 
other M1 transitions are enhanced. The latter can be interpreted as an 
evidence of considerable contributions of quasideuteron configurations to 
the low-spin K=0 states in $^{46}$V.

\section{RESULTS FOR $^{58}$CU}
\begin{figure}[t]
\epsfxsize 15.0cm \centerline{\epsfbox{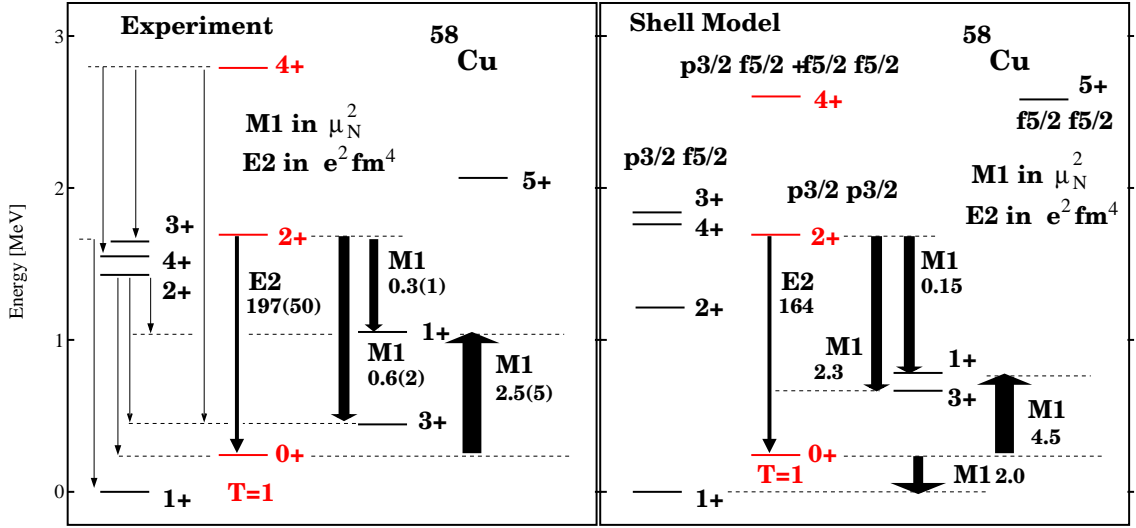}}
\caption{(Left) Part of the low spin level scheme of $^{58}$Co observed in 
$^{58}$Ni\,(p,\,n\, $\gamma$)$^{58}$Cu reaction. Transitions for which
new branching ratios or absolute strengths (with an exception of 
$0^+_1 \rightarrow 1^+_2$ transition) were measured are shown.
(Right) Results of shell model calculations. The main components of the 
wave functions and transition strengths are shown.}
\label{fig:levels}
\end{figure}

In this section we focus on the results of a very recent work of 
I.Schneider {\em et al.} (work in progress) where the low-spin 
structure of the odd-odd N=Z nucleus $^{58}$Cu was investigated up to an 
excitation energy of 4 MeV.
Excited states of $^{58}$Cu were populated in the 
$^{58}$Ni\,(p,\,n$\gamma$)$^{58}$Cu fusion evaporation reaction. 
Single $\gamma$-spectra
and $\gamma\gamma$-co\-in\-ciden\-ce spectra of the depopulating
$\gamma$-cascades in $^{58}$Cu were measured with high energy
resolution. 
Part of the low spin level scheme of $^{58}$Cu 
constructed from the obtained data is shown in Fig.\,\ref{fig:levels}. 
Our new data together with some recent medium and high spin data for 
$^{58}$Cu from Rudolph. D {\em et al.} \cite{Exp1} enrich our knowledge 
of the structure of $^{58}$Cu. 

To get a qualitative understanding of the structure of the low-lying 
states in $^{58}$Cu we have performed simplified spherical shell model 
calculations -- one odd proton and one odd neutron were allowed to 
occupy $p_{3/2}$, $f_{5/2}$ and $p_{1/2}$ orbitals. The doubly closed 
shell nucleus $^{56}$Ni was considered as the inert core. 
The Surface Delta Interaction was used as residual interaction with the same 
strength of pp, nn and pn T=1 interaction ($A_1=0.5$ MeV) and slightly 
weaker strength of pn T=0 interaction ($A_1=0.45$ MeV). The single particle 
energies are taken to be similar to those from \cite{Trache}.      
The result of shell model calculations for the low-lying states are 
shown in Fig.\ref{fig:levels}. 

Isovector M1 transitions are of our special interest. As it follows from 
the quasideuteron scheme the total M1 $0^+_1 \rightarrow 1^+$ transition 
strength for one proton and one neutron in the $p_{3/2}$ orbital amounts to 
13 $\mu_N^2$ \cite{Lis99}. If one uses quenched (by a factor of 0.7) spin 
g-factors this number reduces to 7 $\mu_N^2$. However in the configurational 
space that involves $p_{3/2}$, $f_{5/2}$ and $p_{1/2}$ orbitals one can 
construct five low-lying (below 4 MeV) $J^\pi=1^+$,T=0 states. Therefore one 
can expect significant fragmentation of quasideuteron strength. 
From the previous studies the lifetime of only one $1^+$ state, namely 
the $1^+_2$ state is known: 
$\tau=114(29)$fs. This lifetime corresponds to 
B(M1,$0^+_1 \rightarrow 1_2^+$)= 2.5(5) $\mu_N^2$. This transition is 
predicted by the shell model to be the strongest one among all 
other $0^+_1 \rightarrow 1_i^+$ transitions in $^{58}$Cu. However the shell 
model 
overestimates its strength as well as the strength of 
the $2^+_2 \rightarrow 3_1^+$ transition. This indicates that  stronger 
configuration mixing occurs for the $0^+_1$, $1^+_2$, $2^+_2$ and 
$3_1^+$ states which are predicted by the shell model to contain  
large quasideuteron $[\pi p_{3/2} \times \nu p_{3/2}]$ component. 

In the present experiment we have observed  the $2^+_2 \rightarrow 0_1^+$ 
transition for the first time and have measured its branching ratio.  
Taking into account that the lifetime of the $2^+_2$ level is known we have 
deduced 
the absolute values of the E2 $2^+_2 \rightarrow 0_1^+$, 
M1 $2^+_2 \rightarrow 1_2^+$ and M1 $2^+_2 \rightarrow 3_1^+$ transition 
strengths ( see Fig \ref{fig:levels}). It is interesting to note that 
B(E2,$2^+_2 \rightarrow 0_1^+$) value for $^{58}$Cu is not very different 
from the B(E2,$2^+_1 \rightarrow 0_1^+$) value (130(7) e$^2$fm$^4$) for 
the isospin triplet partner $^{58}$Ni.  We note also that the deduced B(M1) 
values are not very large. It indicates that contribution of the quasideuteron 
configurations to the $2^+_2$ state is smaller than to the $1^+_2$ and 
$3^+_1$ states. This also follows from the structure of the shell model wave 
functions. It would be very interesting to know how the remaining  M1 
quasideuteron strengths are distributed in $^{58}$Cu.

\section{SUMMARY}
 
In summary, using new experimental data obtained in Cologne we have 
illustrated a mechanism of fragmentation of strong isovector M1 transitions 
in deformed and spherical odd-odd N=Z nuclei. The reduction of the M1 
$0^+_1 \rightarrow 1^+_1$ transition strength in deformed $^{46}$V 
is caused by the strong coupling of the quasideuteron configurations to 
the collective core while the suppression of the M1 transitions in near 
spherical $^{58}$Cu nucleus is due to the strong mixing of quasideuteron 
configurations with other non-collective two nucleon configurations.   
The studies reported in the 
present paper indicate that accurate lifetime measurements are needed for 
odd-odd N=Z nuclei to get clearer understanding of their structure.

We thank  J.~Eberth, A.~Gelberg, K.~Jessen, R.~V.~Jolos   
for valuable discussions. We gratefully acknowledge the cooperation with 
the Tokyo group: T.~Otsuka and Y.~Utsuno for the shell model calculations for 
$^{46}$V. This work supported in part by the DFG under 
Contracts no. Br 799/10-1, Br 799/9-3, Pi 393/1-1 and by the US DOE under 
Contract No. DE-FG02-91ER-40609.

\end{document}